\documentclass[11pt]{article}
\usepackage{amsmath}
\usepackage{amsthm}
\usepackage{amsfonts}
\usepackage{amssymb}
\usepackage{color}
\usepackage{graphicx}
\usepackage{url}
\usepackage{subfig}

\graphicspath{{./figures/}}

\newcommand{\gd}{\mathbf{g}}

\usepackage[top=2.5cm, bottom=3.5cm, left=2.5cm, right=2.5cm]{geometry} 
\usepackage[noend]{algorithmic}
\usepackage{algorithm}
\floatname{algorithm}{Algorithm}

\title{$T$-$\Omega$ formulation with higher order hierarchical basis functions for non simply connected conductors}

\author{Ahmed Khebir\\
		EMWorks Inc., Lasalle, Qu\'{e}bec,\\
		H8N 2X2, Canada. 
\and	   
        Pawe{\l} D{\l}otko \\
        Department of Mathematics,\\
        Swansea University, Singleton Park,\\ 
        Swansea SA2 8PP, UK.
  \and
        Bernard Kapidani \\
		Polytechnic Department of Engineering\\
		and Architecture (DPIA),\\
		Universit\`{a} di Udine,\\
		 33100 Udine, Italy.
	\and
		Ammar Kouki \\
		Ecole de Technologie Superieure,\\ 
		Montreal, Qu\'{e}bec, H3C 1K3, Canada.
	\and
	Ruben Specogna \\
		Polytechnic Department of Engineering\\
		and Architecture (DPIA),\\
		Universit\`{a} di Udine,\\
		 33100 Udine, Italy.
}

\date{ \today}

\begin{document}


\maketitle

\begin{abstract}
This paper extends the $T$-$\Omega$ formulation for eddy currents based on higher order hierarchical basis functions so that it can deal with conductors of arbitrary topology. To this aim we supplement the classical hierarchical basis functions with non-local basis functions spanning the first de Rham cohomology group of the insulating region. Such non-local basis functions may be efficiently found in negligible time with the recently introduced D{\l}otko--Specogna (DS) algorithm.
\end{abstract}

\textbf{keywords}: finite element method, eddy currents, high order, hierarchical basis functions, cuts, cohomology.

$T$-$\Omega$ Finite Element formulation for solving eddy currents problems is computationally very attractive given that it reduces the number of unknowns by exploiting the fact that the magnetic field is irrotational in the insulating region. Yet, the standard Finite Element formulation based on Whitney edge elements \cite{ren} and the Discrete Geometric Approach (DGA) counterpart described in \cite{cicp} provide a current density which is only uniform inside each mesh element. Higher order basis functions are therefore very attractive since they yield greater accuracy for a given computational cost and smoother current density vector field.

Among the various possibilities to obtain a high order of convergence, the hierarchical basis functions introduced in \cite{webb2} and \cite{webb4}  are particularly appealing. They allow to have a good control over the distribution of degrees of freedom (dofs) given that different orders can coexist on the same mesh. Moreover, these different orders may be set by hand or, as in $p$-adaptivity, automatically by a mesh refinement scheme.

The $T$-$\Omega$ Finite Element formulation for solving eddy currents problems using hierarchical high order basis functions has been introduced in \cite{webb3}. Yet, the authors of \cite{webb3} assume that the conducting region is simply connected. Since in the engineering practice in most cases this assumption is overly restrictive, the aim of this paper is to fill this gap by proposing a formulation valid for manifold conductors of arbitrary topology. A previous attempt to solve this issue is reported in \cite{he}, even though necessary details for such a solution are missing, see Sec. \ref{cex}.

Let us assume that the conductors form a combinatorial three-manifold $\mathcal{K}_c$ and that $\mathcal{K}=\mathcal{K}_c \cup \mathcal{K}_a$, where $\mathcal{K}_a$ represents the insulating region and $\mathcal{K}$ the whole computational domain. We assume that $\mathcal{K}$ is simply connected since this is the case in the great majority of problems while also making the presentation easier. However, this assumption does not affect the generality of the results of this paper, given that it may be removed as described in \cite{cpc}.

If one wants to solve the eddy current problem with the magnetic scalar potential in configurations containing topologically nontrivial conductors (for example, conductors with ``handles'' like a torus), representatives of first cohomology group generators of insulator are the only objects that make the problem well defined. Thus, the idea proposed in this contribution is to supplement the basis functions introduced in \cite{webb3} with non-local basis functions spanning the \emph{first de Rham cohomology group} $H^1_{dR}(\mathcal{K}_a)$ \cite{edm} of the insulating region.

To achieve good overall performances, it is mandatory to address this topological pre-processing efficiently. With classical algebraic methods like \cite{cmes} based on reduction of the complex $\mathcal{K}_a$ followed by a Smith Normal Form (SNF) computation on the reduced complex, finding such non-local basis functions easily becomes the bottleneck of the whole simulation, see \cite{topo}. This is the main reason why such rigorous methods to solve this issue did not take off in electromagnetic solvers.

In this paper we propose to perform the topological pre-processing with the DS algorithm \cite{cpc}, whose typical computational complexity is linear. Yet, the set of representatives provided by the DS algorithm is a \textit{lazy} cohomology basis \cite{cpc}, \cite{lazy}: the provided set of representatives span the needed cohomology group, but contains additional, dependent elements. The size of the lazy basis is no more than twice the size of a standard cohomology basis and with moderate effort one may produce a standard cohomology basis (see \cite{cpc}) given a lazy one. However, it has been verified that this technique does not provide any speedup in the solution of the electromagnetic problem while giving exactly the same solution in terms of induced currents up to linear solver tolerance.

The paper is organized as follows. In Section II we introduce the novel $T$-$\Omega$ formulation to solve eddy current problems with high order hierarchical basis functions which works also when conductors have arbitrary topology. Moreover, we survey the algorithm used to extract the lazy cohomology basis. Section III presents the results obtained when solving two TEAM problem benchmarks. Finally, in Section IV, the conclusions are drawn.

\section{Novel $T$-$\Omega$ formulation}
\subsection{de Rham cohomology}
\begin{figure}[!t]
\center
  \includegraphics[scale=0.2]{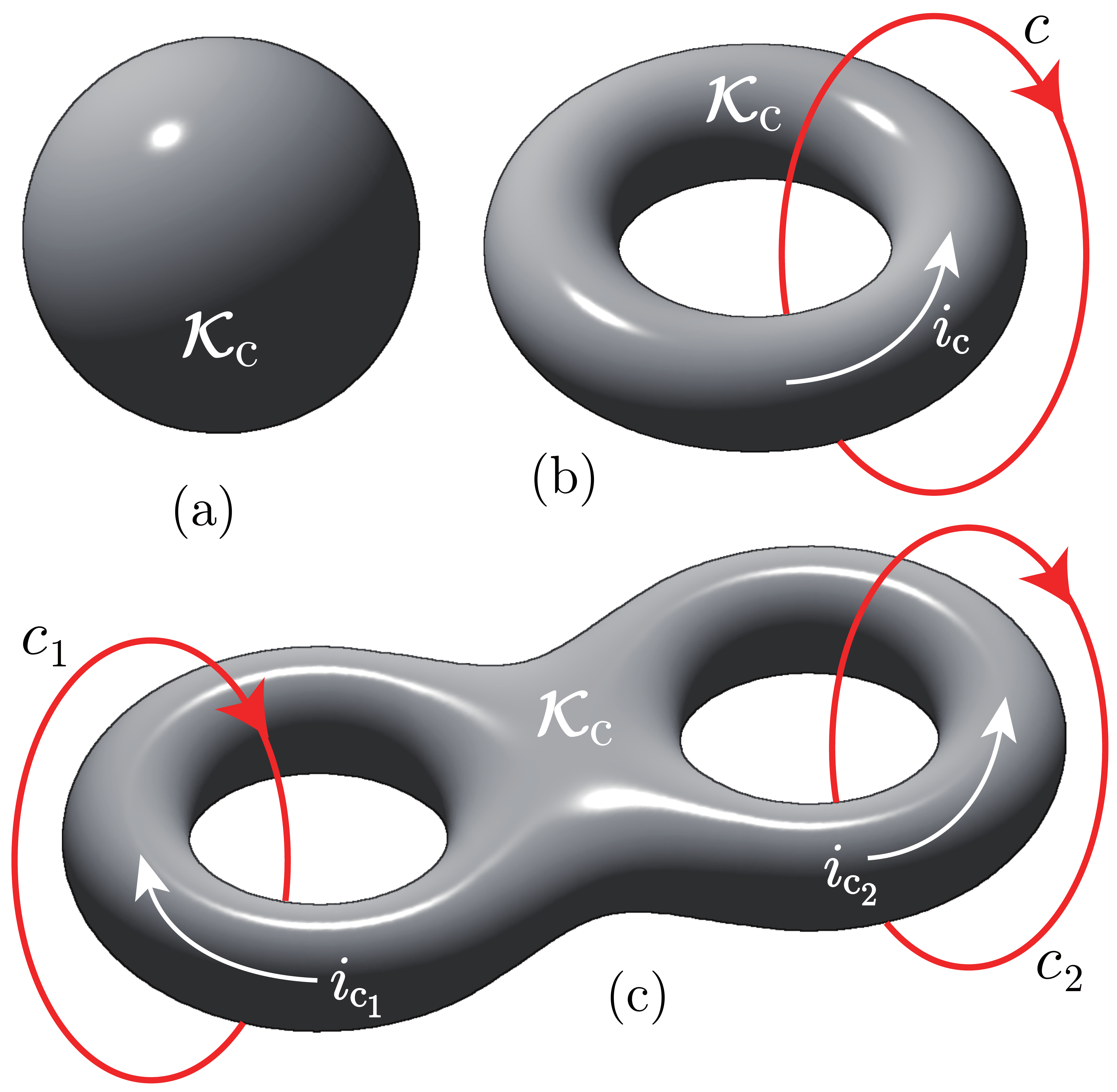}
  \caption{The insulating region $\mathcal{K}_a$ is obtained as the complement of a with respect to a box (not represented in the picture for clarity) of (a) a trivial conductor, (b) a solid torus, (c) a two-fold solid torus, i.e. a conductor with two handles.}
  \label{fig:derham}
\end{figure}
If $\mathcal{K}_a$ is simply-connected (or topologically trivial), as in Fig. \ref{fig:derham}a, a curl-free vector field, like the magnetic field $h$, can be expressed as the gradient of the magnetic scalar potential $\Omega$.

Topology starts playing a role in the $T$-$\Omega$ formulation when $\mathcal{K}_a$ is not simply-connected. Let us consider an example where $\mathcal{K}_a$ is the complement of a conductive torus $\mathcal{K}_c$ with respect to a box as in Fig. \ref{fig:derham}b. The magnetic field $h$ is curl-free in $\mathcal{K}_a$ but its circulation on the loop $c$ of Fig. \ref{fig:derham}b, because of Amp\`{e}re's law, has to match the electric current $i_c$ that flows around $\mathcal{K}_c$.
%
The consequence is that the gradient of the magnetic scalar potential is not enough to span the space where the magnetic field $h$ lives.

In the following we use concepts of algebraic topology that due to the limited space cannot be reproduced here. Please consult~\cite{munkres} for a formal introduction or \cite{sinum},~\cite{cicp},~\cite{cpc} for an informal one.

In algebraic topology, the \emph{first de Rham cohomology group} $H^1_{dR}(\mathcal{K}_a)$ of $\mathcal{K}_a$, is exactly the space of curl-free vector fields in $\mathcal{K}_a$ that are not gradients \cite{edm}. Therefore, by its very definition, $H^1_{dR}(\mathcal{K}_a)$ is the space we have to add to the space generated by the well known hierarchical basis functions described in \cite{webb}. It is known that the dimension of this space is equal to the first Betti number $\beta_1(\mathcal{K}_a)$ of $\mathcal{K}_a$ \cite{edm}.

Focusing again on the example in Fig. \ref{fig:derham}b, $\beta_1(\mathcal{K}_a)=1$, thus the basis for the $H^1_{dR}(\mathcal{K}_a)$ space is composed by just one element called \emph{generator}. Let us construct the generator $g \in H^1_{dR}(\mathcal{K}_a)$ as a curl-free field that has a circulation $1$ on a loop linking the conductor as $c$ in Fig. \ref{fig:derham}b. If one now considers the product $i_c\, g\in H^1_{dR}(\mathcal{K}_a)$, by varying the independent current $i_c$ one is able to span the whole $H^1_{dR}(\mathcal{K}_a)$ space. Therefore, $i_c$ is a new degree of freedom that has to be added as an additional unknown of the eddy current problem.

This property is generalizable to more complicated topologies, see~\cite{massey},~\cite{cpc} for a more formal explanation.
Here we informally show what happens in a more complicated example, i.e. let us consider a two-fold solid torus as $\mathcal{K}_c$, see Fig. \ref{fig:derham}c. Here, $\beta_1(\mathcal{K}_a)=2$ (the conductor has two handles), thus there are two independent currents $i_{c_1}$ and $i_{c_2}$. The $H^1_{dR}(\mathcal{K}_a)$ space is generated by $i_{c_1}\,g_1+i_{c_2}\,g_2$, where the two generators $g_1, g_2$ are such that
\begin{equation}
\int_{c_i} g_j \cdot \hat{t} dl=\delta_{ij},
\end{equation}
$\hat{t}$ is the tangent vector of the loop $c_i$ and $\delta_{ij}$ is the Kronecker delta.

In general, the magnetic field $h$ is represented as
\begin{equation}
h=\nabla \Omega + \sum_{k=1}^{\beta_1(\mathcal{K}_a)} i_{c_k}\,g_k,
\end{equation}
where the space of gradients of the magnetic scalar potential $\Omega$ has been already defined in~\cite{webb} and each independent current has to be included as an additional degree of freedom to span the whole space of the eddy current problem solution.

\subsection{Construction of $H^1_{dR}(\mathcal{K}_a)$ generators}
\begin{figure}[!t]
\center
  \includegraphics[width=\columnwidth]{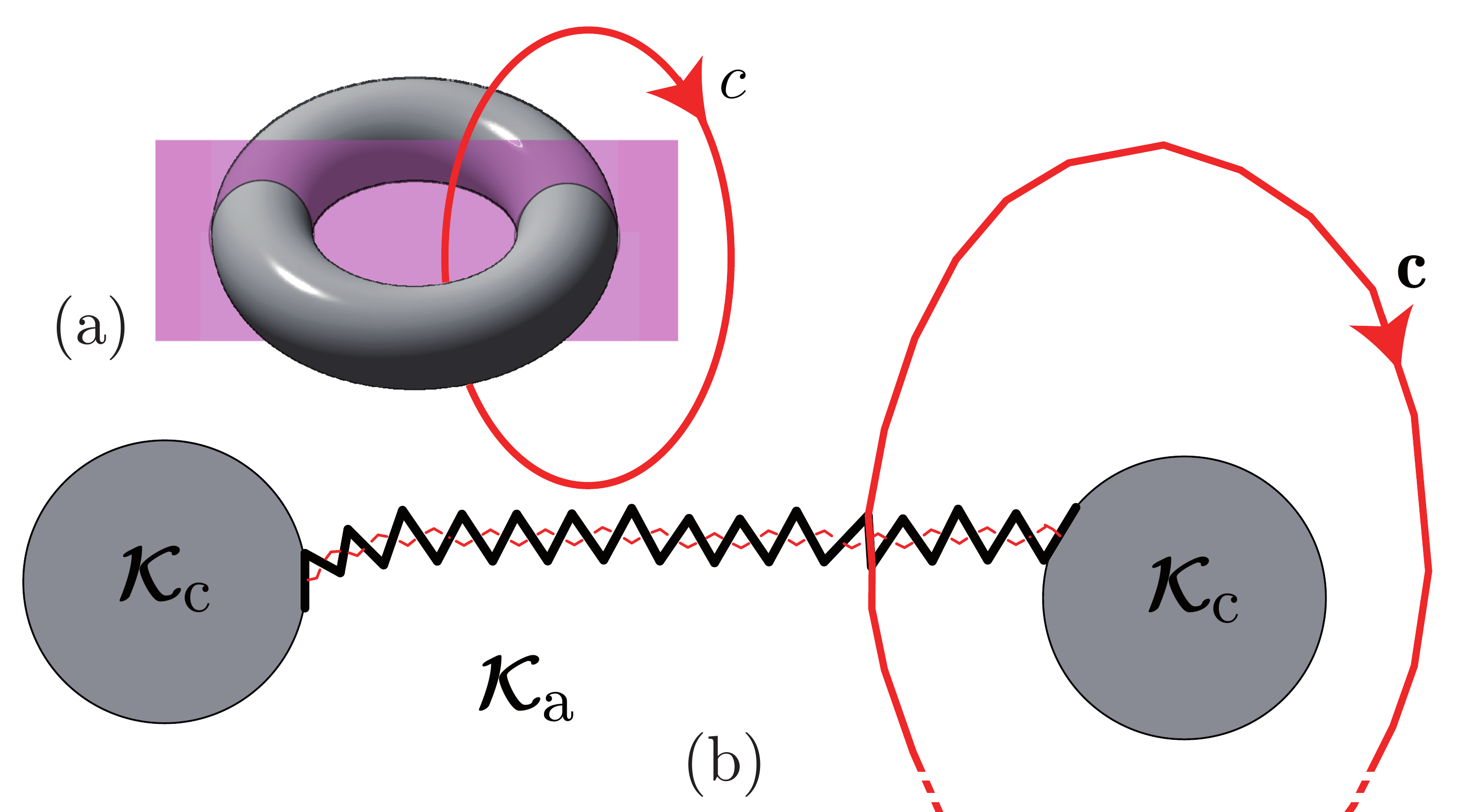}
  \caption{(a) Let us consider a cross section of the example in Fig. \ref{fig:derham}b. (b) The thick black edges belong to the thick cut $\gd$ (i.e. to the support of the representative of the cohomology generator $H^1(\mathcal{K}_a, \mathbb{Z})$). The coefficients assigned to thick edges are such that $\mathbf{C_a} \gd = \mathbf{0}$, where $\mathbf{C_a}$ is the restriction of $\mathbf{C}$ to $\mathcal{K}_a$, and that the circulation on all discrete cycles linking the conductive torus like $\mathbf{c}$ is $1$.}
  \label{fig:thickcut}
\end{figure}
Let us try to replicate the properties of $g$ in the example of Fig. \ref{fig:derham}b at the discrete level. Let us call $\gd$ the discrete counterpart of $g$. First, the circulation of $\gd$ on a loop $\mathbf{c}$ made of edges that links the conductor like $c$ has to be $1$ as for $g$. Therefore, it is natural to define $\gd$ as a discrete field having integer coefficients assigned to mesh edges. Second, we want $\gd$ to be curl-free as its continuous equivalent. This translates in the discrete case to the condition $\mathbf{C_a} \gd = \mathbf{0}$, where $\mathbf{C_a}$ is the face-edge incidence matrix of the mesh restricted to $\mathcal{K}_a$. That is, the discrete circulation of $\gd$ over the edges in the boundary of each face of the mesh must be zero. It is possible to prove that such edge discrete fields may be interpreted as the representatives of the generators of the first cohomology group over integers $H^1(\mathcal{K}_a, \mathbb{Z})$ \cite{massey}, \cite{cpc}. Then, to retrieve $g$, we just need to interpolate $\gd$ with the standard Whitney edge elements basis functions \cite{bossavit}. Thus, it is evident that there is no need to use higher order basis functions to span the de Rham space.

The $H^1(\mathcal{K}_a, \mathbb{Z})$ are realized using the \emph{thick cut} technique \cite{thickcut}, see Fig. \ref{fig:thickcut}.

The advantage in switching to cohomology over the integers is that the generators of $H^1(\mathcal{K}_a, \mathbb{Z})$ can be rigorously and efficiently computed. In this respect, it is important to note that the property of the basis functions of being hierarchical is of fundamental importance: thus, the term of $h$ related to $H^1(\mathcal{K}_a, \mathbb{Z})$, can be accounted for when using higher order by adding it to the scalar first order dofs on the edges of $\mathcal{K}_a$. In what follows we give a detailed description of an algorithm that performs this task.

\subsection{Efficient computation of $H^1(\mathcal{K}_a, \mathbb{Z})$ generators}\label{cex}
\begin{figure}[!t]
\center
  \includegraphics[width=\columnwidth]{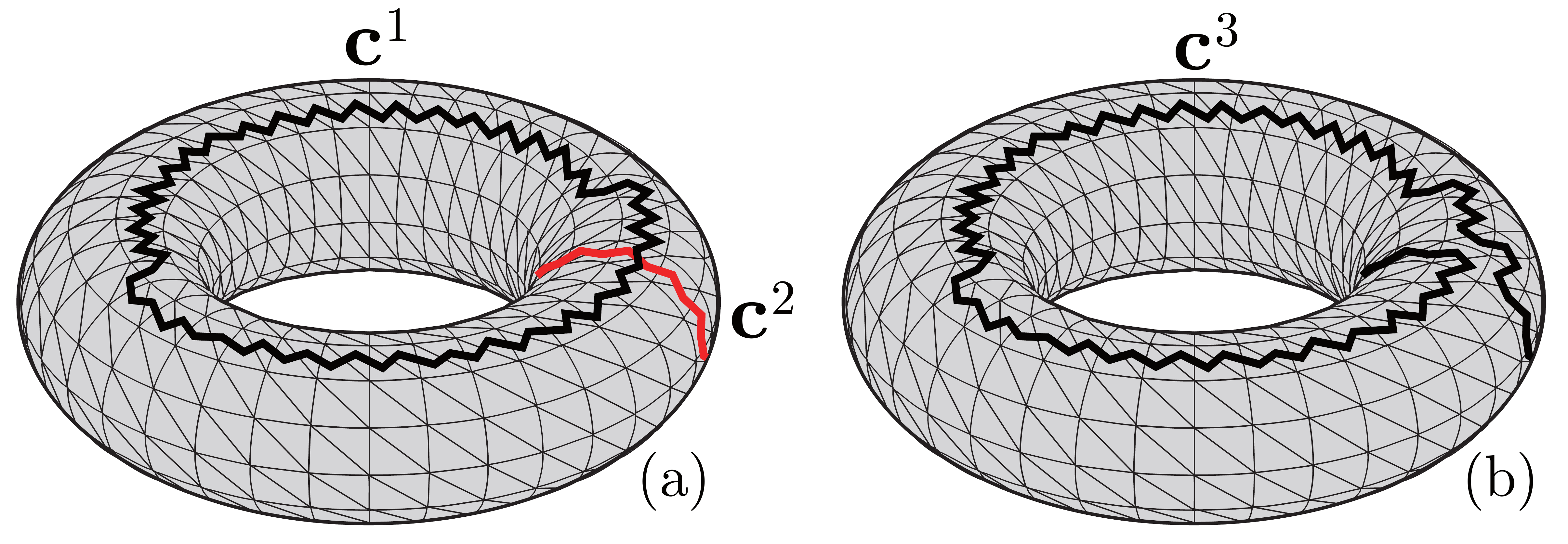}
  \caption{(a) The cycles on the dual complex of $\mathcal{S}$ which represent the supports of the representatives $\mathbf{c}^1$ and $\mathbf{c}^2$ of one of the possible cohomology basis $H^1(\mathcal{S},\mathbb{Z})$. This particular basis is such that the dual cycle of $\mathbf{c}^1$ is bounding in the complement of $\mathcal{K}$, whereas the one of $\mathbf{c}^2$ is bounding inside $\mathcal{K}$. (b) The support of the dual of the representative of a ``mixed'' generator.}
  \label{fig:torusex}
\end{figure}
We find the required non-local basis functions spanning the first de Rham cohomology group space by exploiting the recently introduced idea of \emph{lazy cohomology generators} \cite{cpc}. In what follows we recall the DS algorithm that finds the lazy generators of the first cohomology group $H^1(\mathcal{K}_a,\mathbb{Z})$:
\begin{enumerate}
  \item First the discrete surface $\mathcal{S}=\mathcal{K}_c \cap \mathcal{K}_a$ (see Fig. \ref{fig:torusex}a) is extracted and its first cohomology group generators $H^1(\mathcal{S},\mathbb{Z})$ are computed with a combinatorial algorithm with linear worst-case complexity \cite{cpc}. In Fig. \ref{fig:torusex}a the supports of the dual in $\mathcal{S}$ of two possible representatives $\mathbf{c}^1$ and $\mathbf{c}^2$ of the cohomology generators are shown.
  \item \textit{Thinned currents} are found by pre-multiplying the generators of $\mathcal{S}$ by the incidence matrix $\mathbf{C}_c$ between face and edge pairs \cite{cpc} restricted to $\mathcal{K}_c$. 
  \item Finally, a vectorialized version of the ESTT algorithm \cite{cpc} is run on the whole complex $\mathcal{K}$ for all thinned currents at once. The ESTT algorithm is a general version of the Webb--Forghani iterative algorithm \cite{webb} to obtain a discrete field whose discrete curl has to match the thinned current vectors obtained at the previous step. Its typical complexity is linear, even though the worst-case complexity is cubical. The output of the ESTT restricted to $\mathcal{K}_a$ form the required lazy cohomology generators.
\end{enumerate}

The output lazy generators span the $H^1(\mathcal{K}_a,\mathbb{Z})$ cohomology group, but they do not form a base given that they are dependent. In the example, the generator obtained by processing $\mathbf{c}^1$ span $H^1(\mathcal{K}_a,\mathbb{Z})$, whereas the generator obtained with $\mathbf{c}^2$ is not useful given that the dual of $\mathbf{c}^2$ bounds a discrete surface inside $\mathcal{K}_c$ (therefore it is not topologically interesting but it is used anyway in the final system of equations).

Disentangling boundary generators is far from being trivial especially because it is in general not enough to pick half of the generators to produce the suitable basis. This is because the generators of the basis may be ``mixed'', meaning that one of them is a linear combination of the two as $\mathbf{c}^3$ in Fig. \ref{fig:torusex}c. Thus, in general there are no easier ways to do it than the change of basis described in the appendix of \cite{cpc}.

The novel idea of lazy cohomology generators \cite{cpc} is exactly that all boundary generators produce a cut, even if half of them may be eliminated. This approach does not provide a sensible slowdown in the system solver time, whereas greatly speedup the topological pre-processing part. Concerning the quality of the solution, the results with a standard vs lazy basis in terms of eddy currents are the same up to solver tolerance.

This technique is appealing first of all because the topological pre-processing requires mere seconds even on very complicated problems. Moreover, the DS algorithm is proved to be general, therefore it works for every possible input, no matter how complicated the geometry or the topology of the insulating region is.

Finally, we note that the technique proposed by Ren \cite{ren} and used later by He et al \cite{he} lack a number of decisive details. The description of the algorithm does not specify if surface cuts (i.e. boundary generators) are disentangled or not (and, if they are then how). It is also not clear whether the surface cuts (i.e. cohomology generators) are extended in $\mathcal{K}_a$ or in the whole $\mathcal{K}$. The idea of extending topologically trivial sets that seems to be at the root of the algorithm in \cite{ren} suffers from very serious drawbacks that may prevent successful termination of the algorithm. However, given the level of generality of the presentation in \cite{ren}, it is not possible to assess its correctness nor to implement and use it.

\section{Numerical results}
The DS algorithm together with the $T$-$\Omega$ formulation with hierarchical basis functions have been implemented inside the EMS solver (\verb"www.emworks.com"). We validated the software and assessed the performances of first and second order approximation in solving various eddy current problems.

\begin{figure}
  \centering
  \subfloat[1st order]{\label{fig:team71st}\includegraphics[width=0.49\columnwidth]{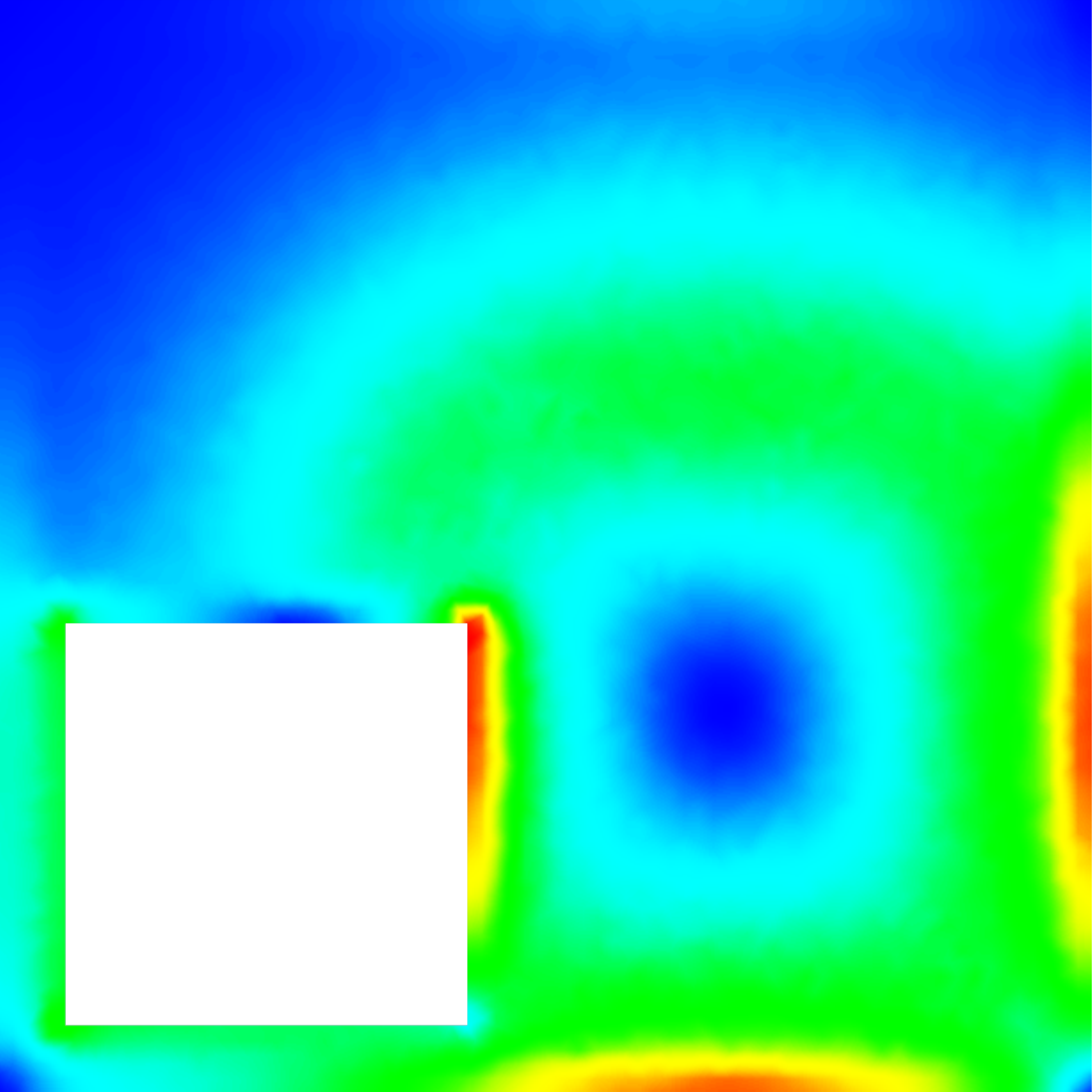}}
  \hfill
  \subfloat[2nd order]{\label{fig:team72nd}\includegraphics[width=0.49\columnwidth]{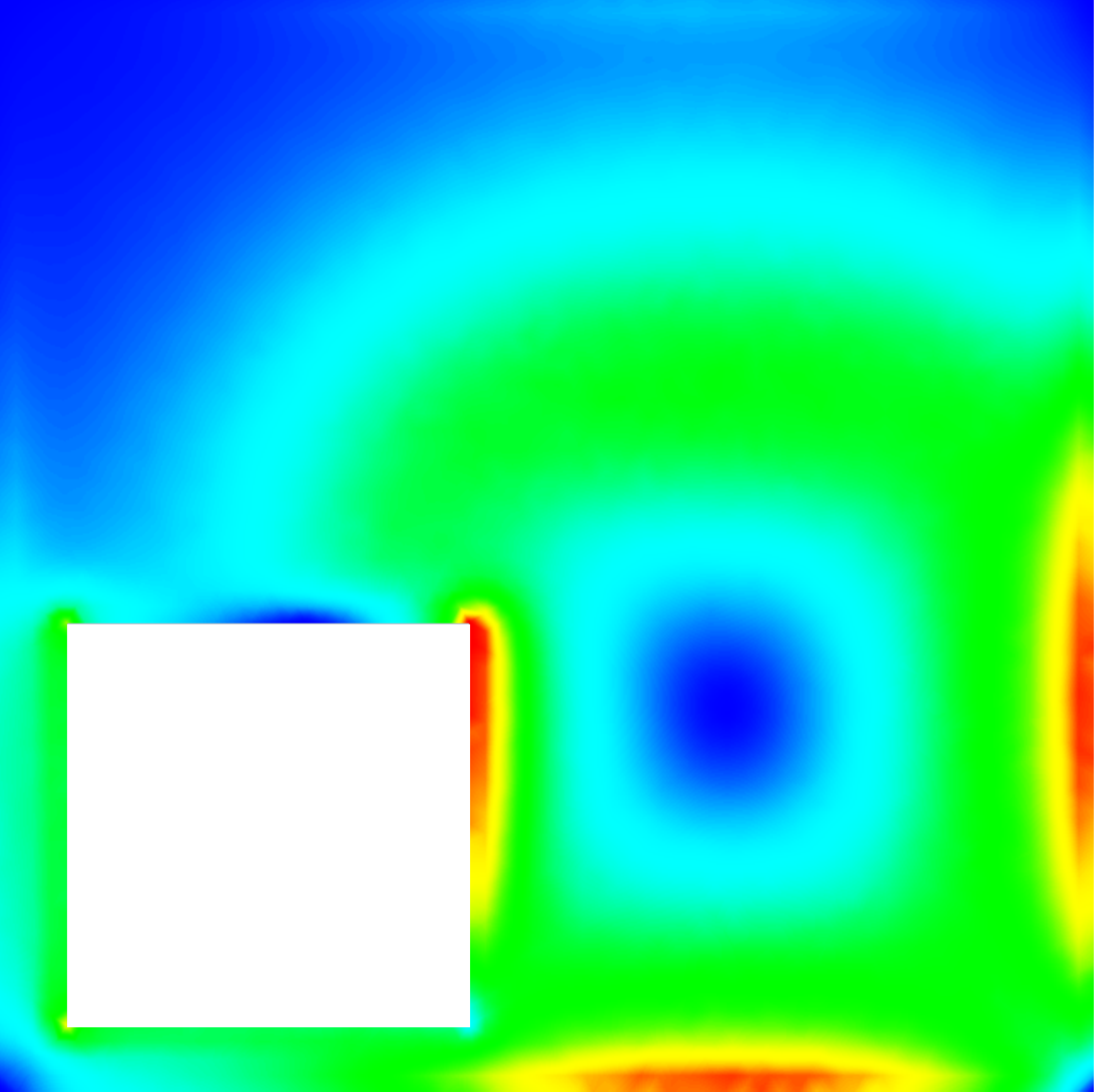}}
  \caption{\label{fig:team71-72}Eddy currents obtained with the electromagnetic code EMS.}
\end{figure}

As a first benchmark, we use the TEAM Workshop problem 7, see \cite{team7}. 
It consists of a racetrack shaped coil 
driven by a time harmonic current (amplitude $2742$AT, frequency
$f = $200Hz) over a square aluminum plate. 
Fig. \ref{fig:team71-72} represents the eddy currents resulting from the solution with first and second order. Table \ref{tab7} contains the comparison of the computed heat losses $P=\int_{\mathcal{K}} \rho j^2$, where $j$ is the current density vector field and $\rho$ is the resistivity.
\begin{table}
\centering
\caption{TEAM problem 7 benchmark}
\label{tab7}
\begin{tabular}{|l|l|l|}
\hline
Number of tetrahedra & Losses 1st order [W] & Losses 2nd order [W] \\ \hline
52$\,$539	&7.11	&8.02\\\hline
126$\,$615	&7.59	&8.83\\\hline
225$\,$478	&9.02	&9.37\\\hline
423$\,$841	&9.32	&9.60\\\hline
755$\,$333  &9.66   &-\\\hline
\end{tabular}
\end{table}

\begin{figure}[!t]
\centering
  \includegraphics[width=0.9\columnwidth]{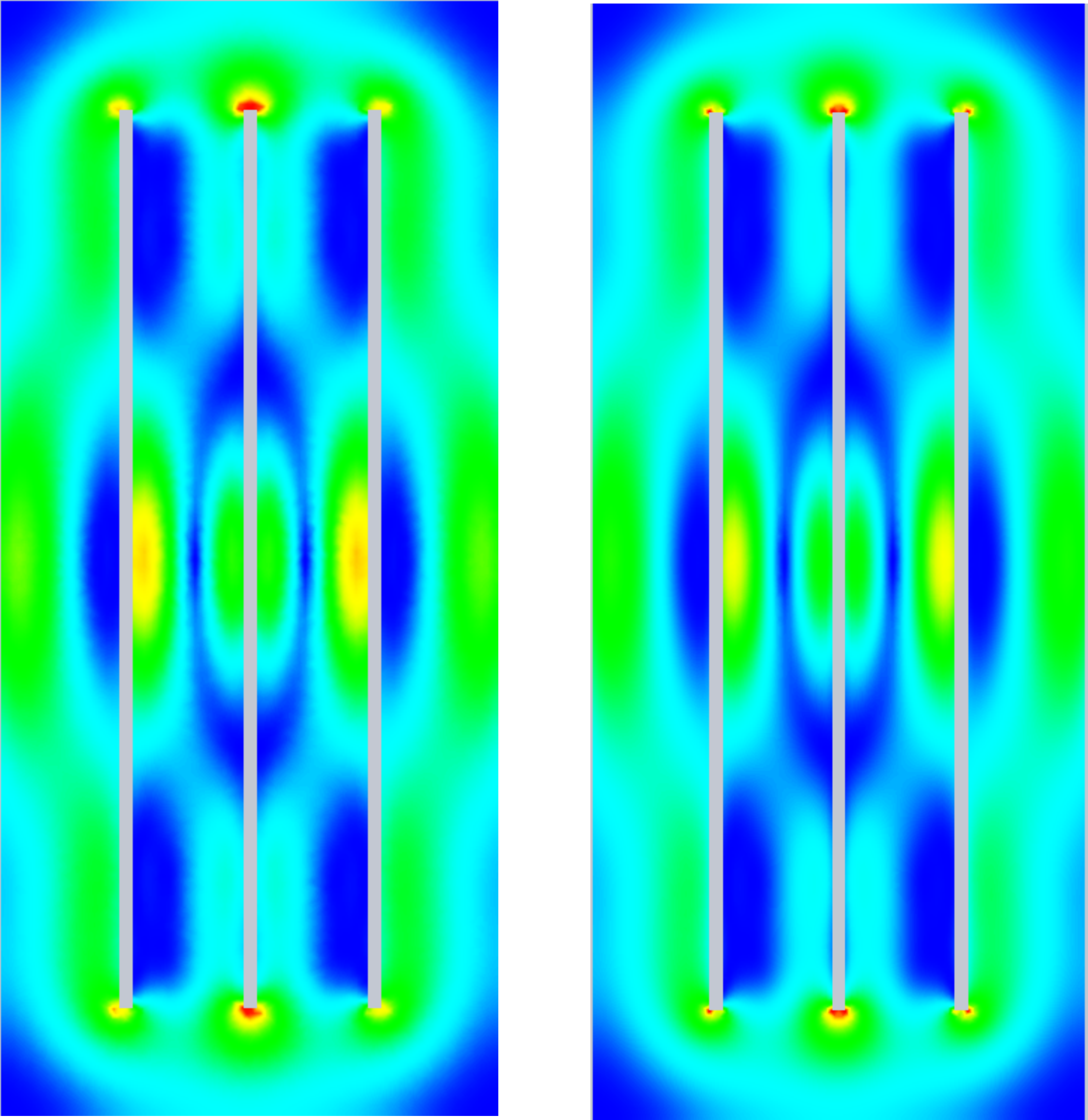}
  \caption{Eddy currents obtained with the electromagnetic code EMS (left, 1st order, right 2nd order) on the TEAM problem 21a-3.\label{fig:team21a-3}}
\end{figure}

As a second benchmark, we consider the TEAM Workshop problem 21, see \cite{team21}. Fig \ref{fig:team21a-3} contains the eddy currents resulting from the solution with first and second order in case of benchmark 21a-3. Table \ref{tab21} contains the comparison of the computed heat losses for the cases 21a-1, 21a-2 and 21a-3.
\begin{table}
\centering
\caption{Loss in Watt for the TEAM problem 21 benchmark}
\label{tab21}
\begin{tabular}{|l|l|l|l|l|}
\hline
Problem & Tetrahedra & Loss 1st ord. & Loss 2nd ord. & Measured\\ \hline
21a-1 & 45$\,$177	&3.19	&3.40 & 3.40\\\hline
21a-2 & 29$\,$577	&1.58	&1.66 & 1.68\\\hline
21a-3 & 199$\,$781	&0.95	&1.04 & 1.25\\\hline
\end{tabular}
\end{table}


\section{Conclusions}
Lazy cohomology generators technology has been integrated inside an eddy current solver employing second order hierarchical basis functions. As expected, the second order formulation indeed provides more accurate results than its first order counterpart. The inclusion of automatic mesh adaptivity to render the second order more efficient is left for further studies.

\end{document}